\documentclass[a4paper]{jpconf}
\usepackage{graphicx}
\begin{document}
\title{First Results from Photon Multiplicity Detector at RHIC }

\author{B. Mohanty (for STAR Collaboration\footnote{Complete author list can be found at the end of this proceeding})}
\address{Variable Energy Cyclotron Centre, 1/AF, Bidhan Nagar, Kolkata - 700064}

\ead{bmohanty@veccal.ernet.in}

\begin{abstract}
We present the first measurement of multiplicity and pseudorapidity
distributions of photons in the pseudorapidity region 
2.3 $\le$ $\eta$  $\le$ 3.7 for different centralities in Au + Au 
collisions at $\sqrt{s_{\mathrm {NN}}}$ = 62.4 GeV.
The pseudorapidity distribution of photons, dominated by
neutral pion decays, has been compared to those of identified 
charged pions, photons, and inclusive charged particles from heavy 
ion and nucleon-nucleon collisions at various energies. 
Scaling of photon yield with number of participating nucleons
and limiting fragmentation scenario for inclusive photon production
has been studied.
\end{abstract}

\section{Introduction}

Inclusive charged particle multiplicity measurements at RHIC have so far 
revealed a lot of information on the nature and dynamics 
of particle production in heavy ion collisions~\cite{brahms,phobos}. 
At midrapidity a significant increase in charged particle production
 normalized to the number of participating nucleons ($N_{\mathrm part}$) 
has been observed for central Au+Au collisions compared to peripheral 
Au+Au and p+p collisions~\cite{brahms,phobos,whitepapers}.  
This has been attributed to the onset of hard scattering processes, 
which scale with the number of binary collisions. Alternatively, in 
the Color Glass Condensate~\cite{cgc} picture  of particle production 
at midrapidity, the centrality dependence could reflect increasing 
gluon density  due to the decrease in the effective strong coupling constant.
Limiting fragmentation (LF)~\cite{limiting_frag} behaviour of
inclusive charged particles have been studied at RHIC. 
It has been observed that inclusive charged particles follow a 
energy independent and centrality dependent limiting fragmentation 
scenario~\cite{brahms,phobos}. It will be interesting to study the 
above physics aspects with inclusive photon multiplicity 
measurements~\cite{photon} at RHIC.
In this paper we present the first photon multiplicity measurements
in the forward rapidity at RHIC for the Au + Au collisions at 
$\sqrt{s_{\mathrm {NN}}}$ = 62.4 GeV in the STAR experiment.

\section{Experiment and data analysis}
The photon multiplicity measurements were done in the 
STAR experiment~\cite{star_nim} at RHIC by a highly granular gas 
based photon  multiplicity detector (PMD) in pseudorapidity 
region 2.3 $\le$ $\eta$  $\le$ 3.7. The details of the construction and design 
of PMD can be found in the following Ref.~\cite{starpmd_nim}.
The data presented here corresponds to 0 to 80\% of Au+Au hadron cross section.
The minimum bias trigger was obtained from the information of the 
following trigger detectors. An array of scintillator
slats arranged in a barrel surrounding the Time Projection chamber (TPC), 
called central trigger barrel (CTB), which measures charged particles and 
two zero degree hadronic calorimeters (ZDCs) at $\pm$ 18 m from the detector
center~\cite{trigger}.The charged particles from TPC within 
$\mid\eta\mid$ $<$ 0.5  was used for the centrality selection for data 
and simulation in the present paper.

\begin{figure}[h]
\begin{minipage}{18pc}
\includegraphics[width=17pc]{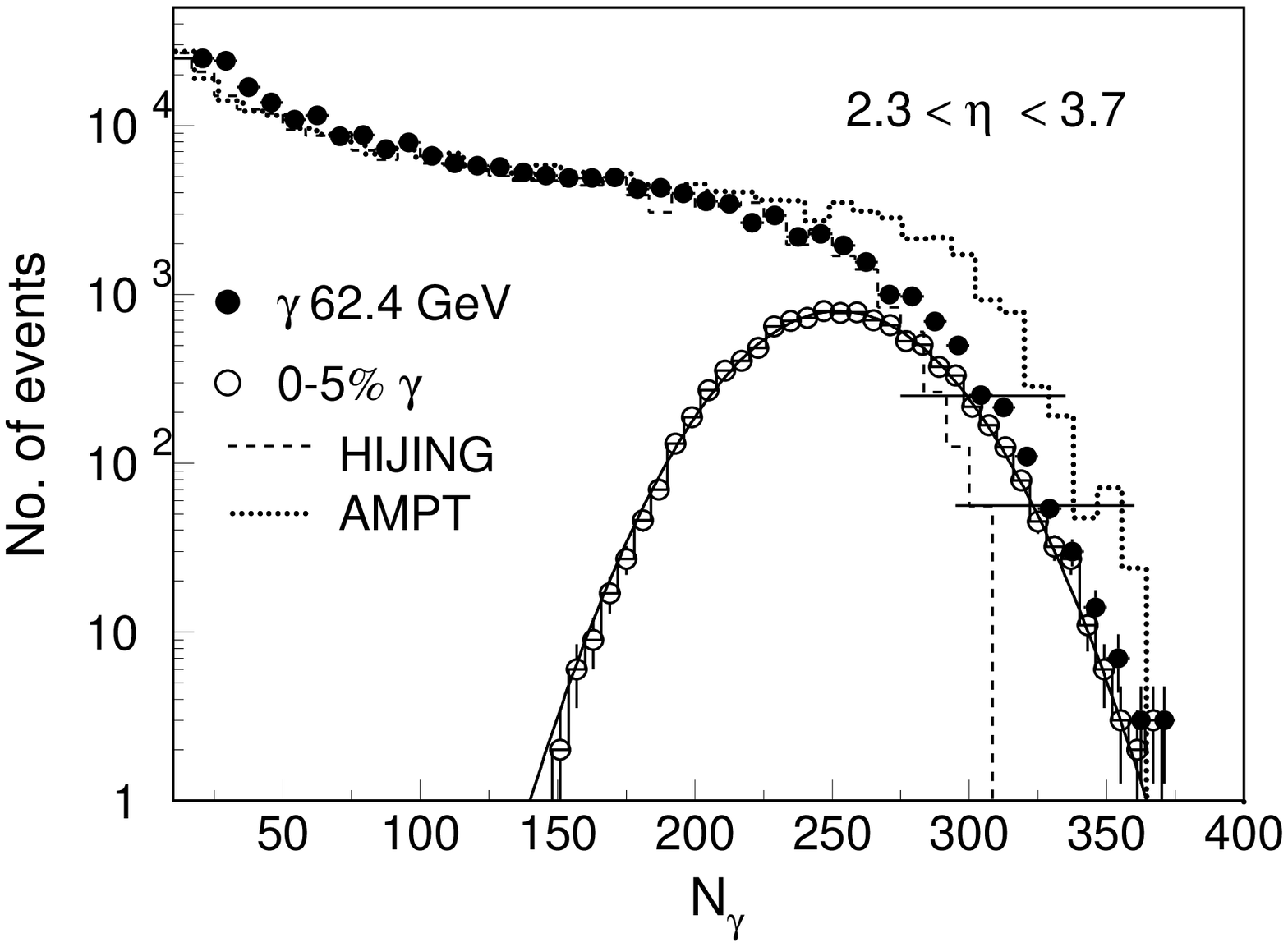}
\caption{\label{fig1}Minimum bias $N_{\mathrm \gamma}$ distribution. 
Comparison with HIJING and AMPT models are shown. 
The solid curve is the fit to a Gaussian.}
\end{minipage}\hspace{2pc}%
\begin{minipage}{18pc}
\includegraphics[width=17pc]{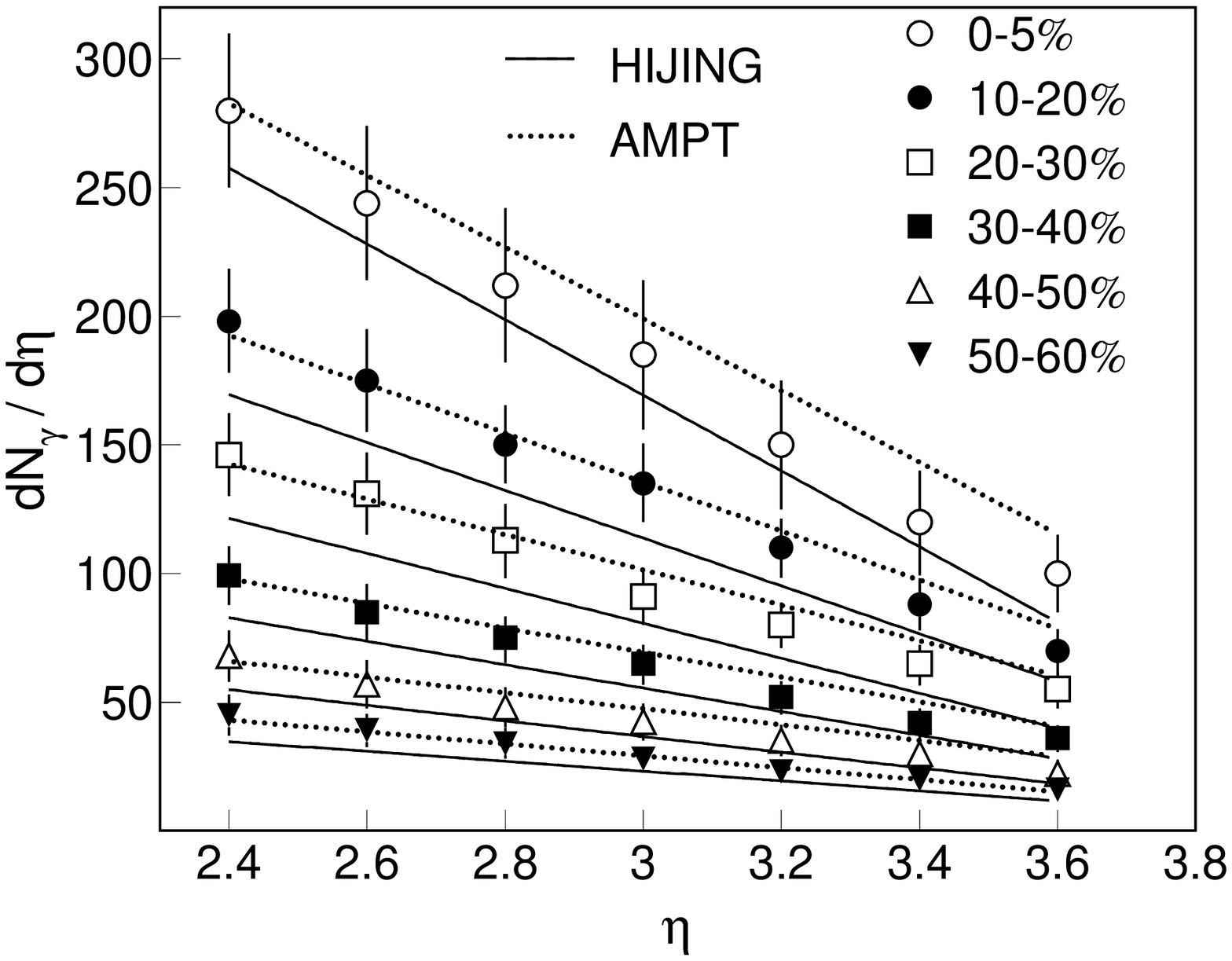}
\caption{\label{fig2}$\frac{d N_{\mathrm \gamma}}{d \eta}$ for various event centrality classes
compared to HIJING and AMPT model calculations.}
\end{minipage}
\end{figure}

In the present analysis, only the data from the preshower plane of the PMD 
has been used. The data analysis proceeded through
the following steps: (a) Calibration of gain of all cells of PMD,
(b) clustering of hits on PMD and (c) photon-hadron discrimination.
Details of each of the above steps of analysis of can be 
found in Ref.~\cite{photon}.

\section{Results}
\subsection{Multiplicity and Pseudorpatidity distributions}
Fig.~\ref{fig1} shows the minimum bias distribution of 
$N_{\mathrm \gamma}$ along with results from HIJING~\cite{hijing} and 
AMPT~\cite{ampt} models. 
We observe that HIJING underpredicts the measured photon multiplicity.
AMPT slightly overpredicts the total measured photon multiplicity for 
central collisions. However, within the systematic errors~\cite{photon} 
it is difficult to make a firm conclusion. The top $5\%$ central 
multiplicity distribution (open circles) is fitted to a Gaussian with
a mean of 252.

\begin{figure}[h]
\begin{minipage}{18pc}
\includegraphics[width=17pc]{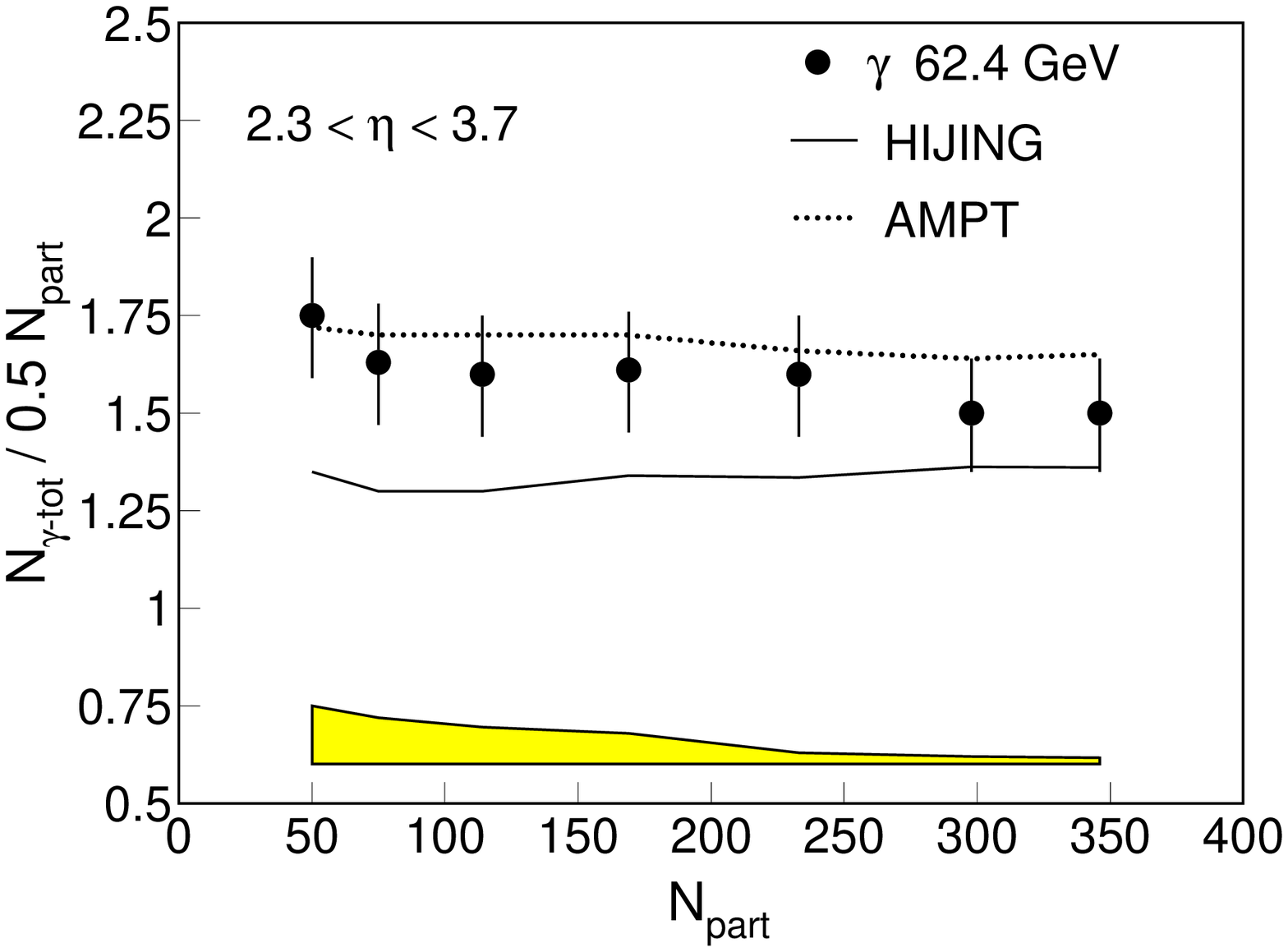}
\caption{\label{fig3}Variation of $N_{\mathrm \gamma}$ 
per participant pair in PMD coverage (2.3 $\le$ $\eta$  $\le$ 3.7) 
as a function of $N_{\mathrm part}$. The error bars shown are 
systematic errors and the lower band reflects uncertainties in 
$N_{\mathrm {part}}$ calculations.}
\end{minipage}\hspace{2pc}%
\begin{minipage}{18pc}
\includegraphics[width=15pc]{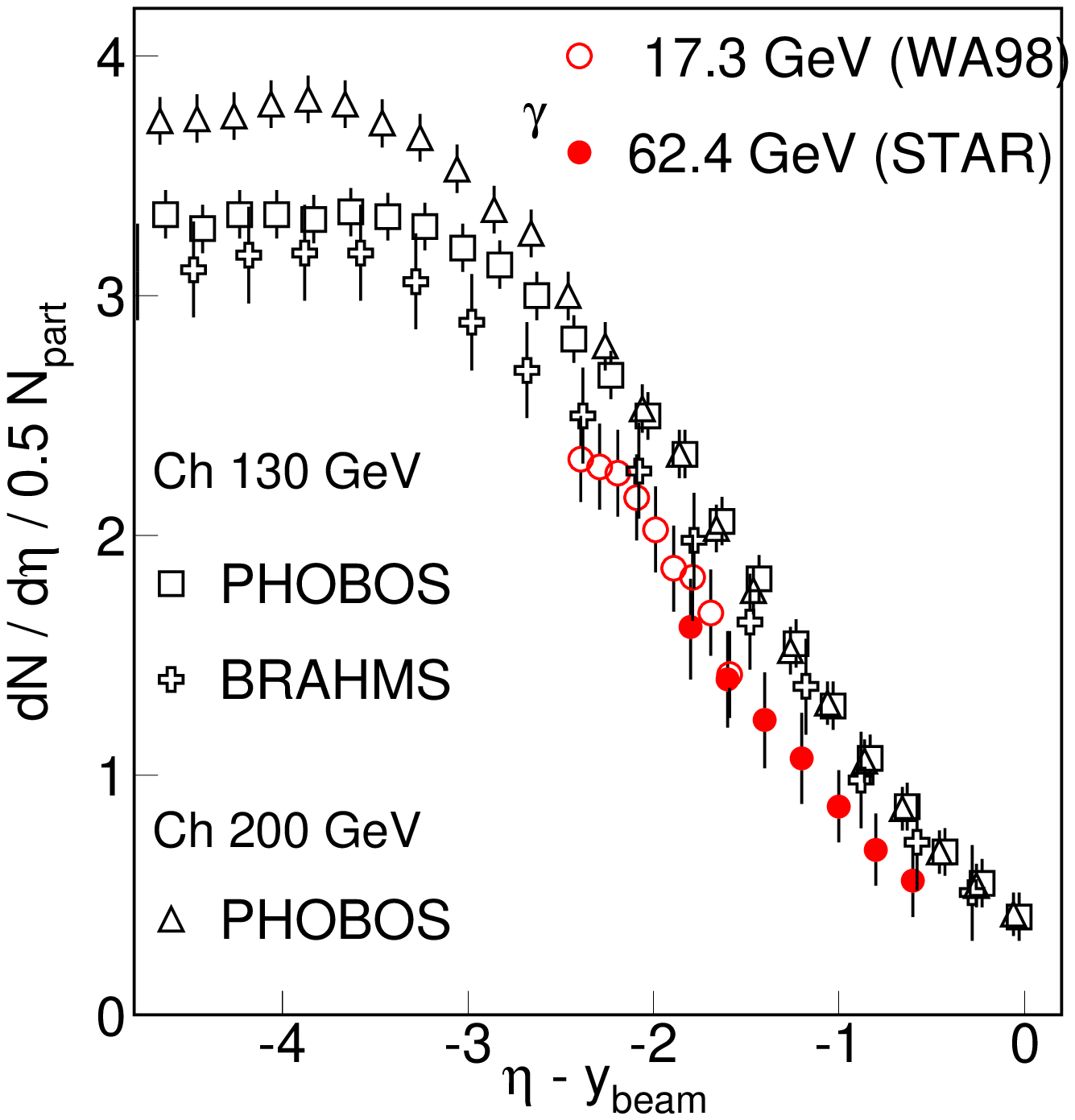}
\caption{\label{fig4}Variation of $\frac{d N_{\mathrm \gamma}}{d \eta}$  and 
$\frac{d N_{\mathrm {ch}}}{d \eta}$ normalized to $N_{\mathrm {part}}$ with $\eta$ - y$_{\mathrm {beam}}$ for central collisions at various collision energies.
The error bars shown are systematic errors.}
\end{minipage}
\end{figure}

Fig.~\ref{fig2} shows the pseudorapidity distribution of photons
for various event centrality classes. 
The errors shown are a quadratic sum of the 
 systematic and statistical errors.
The results from HIJING are systematically 
lower compared to data for mid-central and peripheral events. 
The results from AMPT compare well to the data.

\subsection{Scaling of photon multiplicity with number of participating nucleons}
Fig.~\ref{fig3} shows the variation of total number of 
photons per participant pair in the PMD coverage as a 
function of the number of participants. 
$N_{\mathrm {part}}$ is 
obtained from Glauber calculations~\cite{star_glauber}.
We observe that the total number of photons per participant pair is 
approximately constant with centrality.
The values from HIJING are lower compared to the data. The values from AMPT 
agree fairly well with those obtained from the data.

\subsection{Limiting fragmentation scenario for photons}
Fig.~\ref{fig4} compares the photon spectra in Au + Au collisions 
at $\sqrt{s_{\mathrm {NN}}}$ = 62.4 GeV, with the top SPS energy  photon data for 
Pb + Pb collisions~\cite{wa98_dndy} and charged particle data 
at  $\sqrt{s_{\mathrm {NN}}}$ = 130 and 200 GeV~\cite{brahms,phobos} for Au+Au 
as a function of $\eta$ - y$_{\mathrm {beam}}$ for central collisions. 
The SPS and RHIC photon results are consistent with each other, 
suggesting that photon production follows the LF behavior.
However, the photon multiplicity values are lower compared to 
charged particles.

\begin{figure}[h]
\begin{minipage}{18pc}
\includegraphics[width=16pc]{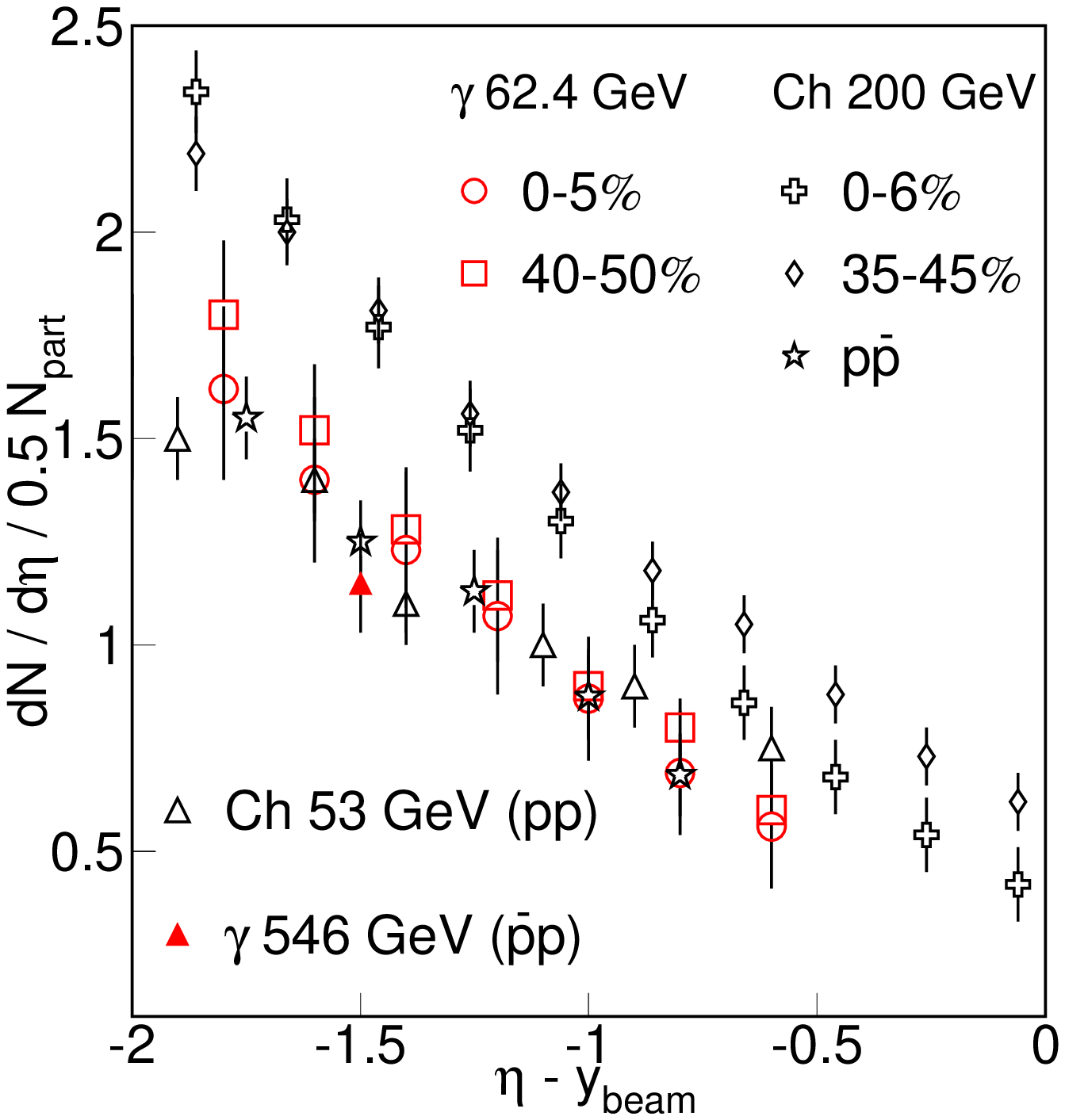}
\caption{\label{fig5}Centrality dependence of limiting fragmentation for 
inclusive photons and charged particles. Comparison to $p \bar{p}$ 
and $pp$ collisions. The error bars shown are systematic errors.}
\end{minipage}\hspace{2pc}%
\begin{minipage}{18pc}
\includegraphics[width=16pc]{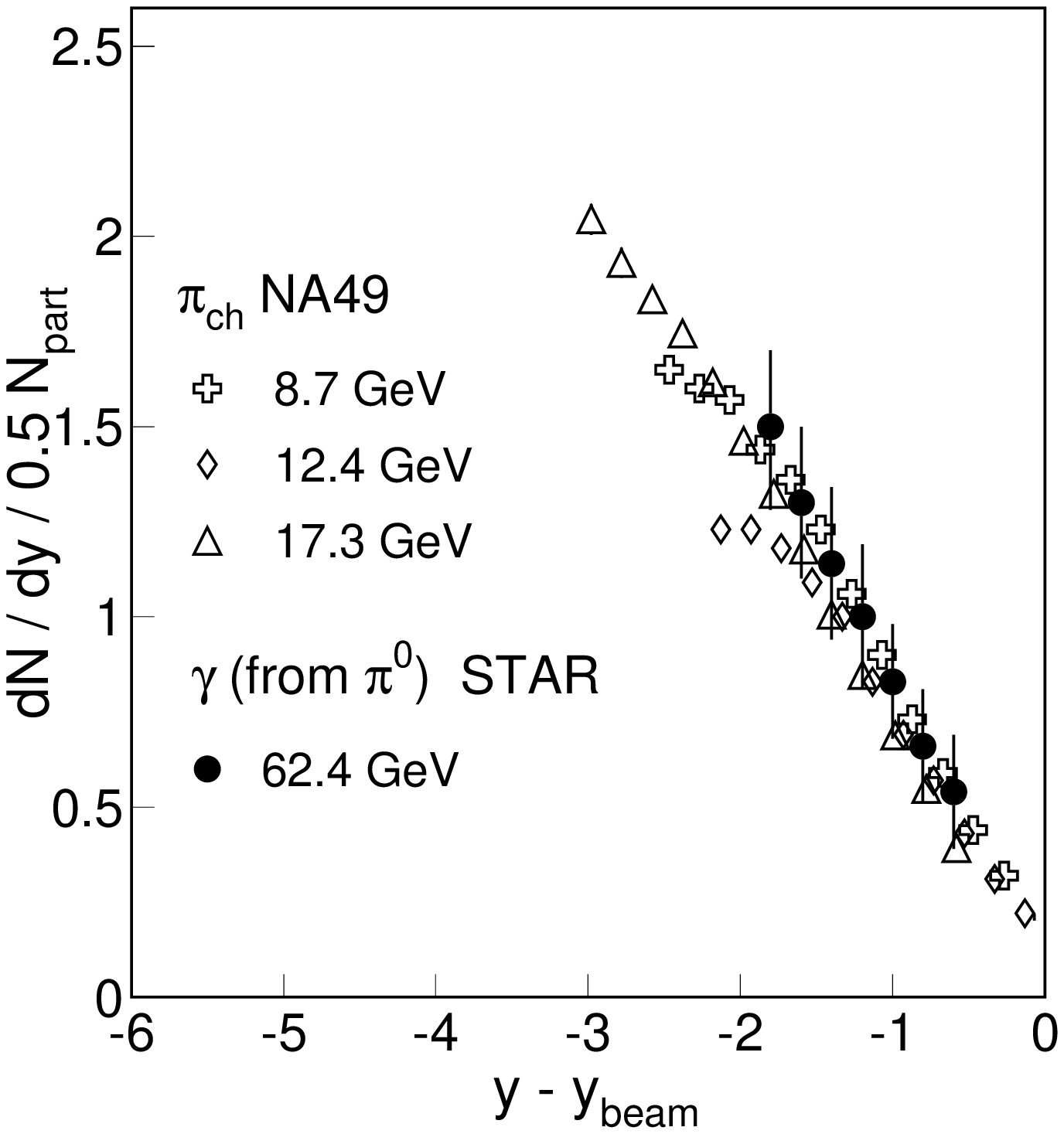}
\caption{\label{fig6}Variation of $\frac{d N_{\mathrm \gamma}}{d \eta}$ scaled down to reflect $\pi^0$ contribution and 
$\frac{d N_{\mathrm {\pi^\pm}}}{dy}$ normalized to  $N_{\mathrm {part}}$  
for central collisions at various collision energies with 
y - y$_{\mathrm {beam}}$. The error bars include the systematic errors.}
\end{minipage}
\end{figure}
In Fig.~\ref{fig5} we study the centrality dependence of the LF behavior 
for inclusive photons and compare the photon spectra with those for charged 
particles. In the forward $\eta$ region, the photon production 
cross section as a function of $\eta$ - y$_{\mathrm {beam}}$ is 
independent of centrality. The dependence of LF spectra on the 
collision system is established by 
the comparison of charged particles spectra from $pp$ and $p \bar{p}$
collisions at $\sqrt{s_{\mathrm {NN}}}$ = 53 and 200 GeV, respectively, 
and $\gamma$ at 546 GeV~\cite{ua5}. We observe that the
photon results in the forward rapidity region from $p \bar{p}$ 
collisions at $\sqrt{s_{\mathrm {NN}}}$ = 546 GeV are
in close agreement with the measured photon yield in 
Au+Au collisions at $\sqrt{s_{\mathrm {NN}}}$ = 62.4 GeV within the LF scenario.

 Fig.~\ref{fig6} shows the total charged pion rapidity density from
 SPS~\cite{na49_pion} and scaled photon
 rapidity density at $\sqrt{s_{\mathrm {NN}}}$ = 62.4 GeV 
 as a function of y-y$_{\mathrm {beam}}$. HIJING calculations
 indicate that about 93-96\% of the photons are from $\pi^0$
 decays.  The photon results in Fig.~\ref{fig6} have been scaled down
 accordingly to reflect approximately twice the $\pi^{0}$ spectrum.
 The results show that pion production in heavy ion collisions in 
 the fragmentation region agrees with the LF picture.

\section{Summary} 
In summary, we have presented the first results of photon multiplicity
measurements at RHIC in the pseudorapidity region 2.3 $\le$ $\eta$  $\le$ 3.7. 
The pseudorapidity distributions of photons
have been obtained for various centrality classes. 
Photon production per participant pair is 
found to be approximately independent of centrality in this 
pseudorapidity region.
Comparison with photon and charged pion data at RHIC and SPS 
energies shows, for the first time in heavy ion collisions, 
that photons and pions follow an energy 
independent limiting fragmentation behavior, as has been previously observed 
for inclusive charged particles. However photons, unlike charged particles, 
follow a centrality independent limiting fragmentation scenario.

\subsection{Acknowledgments}
We thank the RHIC Operations Group and RCF at BNL, and the
NERSC Center at LBNL for their support. This work was supported
in part by the HENP Divisions of the Office of Science of the U.S.
DOE; the U.S. NSF; the BMBF of Germany; IN2P3, RA, RPL, and
EMN of France; EPSRC of the United Kingdom; FAPESP of Brazil;
the Russian Ministry of Science and Technology; the Ministry of
Education and the NNSFC of China; SFOM of the Czech Republic,
FOM and UU of the Netherlands,
DAE, DST, and CSIR of the Government of India; the Swiss NSF.
We acknowledge the help of CERN for use of GASSIPLEX chips in the PMD readout.

\section{References}


\medskip
\begin{thebibliography}{9}

\bibitem{brahms}
               I.G.~Bearden et al., (BRAHMS Collaboration),
               2001 {\it Phys. Lett.} B {\bf 523} 227; 
               2002 {\it Phys. Rev. Lett.} {\bf 88} 202301;
               I. Arsene et al., nucl-ex/0410020.
\bibitem{phobos} 
               B.B.~Back et al., (PHOBOS Collaboration),
               2001 {\it Phys. Rev. Lett.} {\bf 87} 102303;
               2003 {\it Phys. Rev. Lett.} {\bf 91} 052303;
               nucl-ex/0410022.

\bibitem{whitepapers} K. Adcox, et al.,(PHENIX Collaboration),
                      nucl-ex/0410003;
                     J. Adams, et al.,(STAR Collaboration),
                     nucl-ex/0501009.
                     
\bibitem{cgc}
               D. Kharzeev and Marzia Nardi, 
               Phys.Lett. B 507, 121, (2001);
               E. Iancu, A. Leonidov and L. McLerran, 
               hep-ph/0202270.


\bibitem{limiting_frag} 
               J.~Benecke,  T.T. Chou, Chen-Ning Yang and  E. Yen, 
               1969 {\it Phys. Rev.} {\bf 188} 2159;
	        R. Beckmann, S. Raha, N. Stelte and R.M. Weiner,
                1981 {\it Phys. Lett.} B {\bf 105} 411.

\bibitem{photon}
               J. Adams et al.,(STAR Collaboration),
               2005 {\it Phys. Rev. Lett.} {\bf 95} 062301.

\bibitem{star_nim} K. H. Ackermann, et al., 
               2003 {\it Nucl. Instr. Meth.} A {\bf 499} 624.


\bibitem{starpmd_nim} 
               M.M. Aggarwal et al., 
               2003 {\it Nucl. Instr. Meth.} A {\bf 499} 751;
               2002 {\it Nucl. Instr. Meth.} A {\bf 488} 131.


\bibitem{trigger} 
               F.S. Bieser et al., 
               2003 {\it Nucl. Instrum. Meth.} A {\bf 499} 766.

\bibitem{hijing} 
               X-N. Wang and M.~Gyulassy, 
               1991 {\it Phys. Rev.} D  {\bf 44} 3501.

\bibitem{ampt} 
               B. Zhang, C.M. Ko, Bao-An Li and Zi-wei Lin,
               2000 {\it Phys. Rev.} C {\bf 61} 067901.


\bibitem{star_glauber} J.~Adams et al. (STAR Collaboration), nucl-ex/0311017.


\bibitem{wa98_dndy}
               M.M.~Aggarwal et al., (WA98 Collaboration),
               1999 {\it Phys. Lett.} B {\bf 458} 422.

\bibitem{ua5} 
               K.~Alpgard et al., (UA5 Collaboration),
               1982 {\it Phys. Lett.} B {\bf 115} 71; 
               G.J. Alner et al.,
               1986 {\it Z. Phys.} C {\bf 33} 1. 

\bibitem{na49_pion} 
               S.V.~Afanasiev et al., (NA49 Collaboration), 
               2002 {\it Phys. Rev.} C {\bf 66} 054902.


\end{thebibliography}
\end{document}